\documentclass{svmult}
\usepackage[english]{babel}
\usepackage{cite}
\usepackage{amssymb,amsmath}
\usepackage{graphicx,subfigure}
\usepackage{float}
\usepackage[bottom]{footmisc}

\newcommand{\pref}[1]{(\ref{#1})}
\newcommand{\Sec}[1]{Sec.~\ref{#1}}
\newcommand{\Eq}[1]{Eq.~(\ref{#1})}
\newcommand{\Fig}[1]{Fig.~\ref{#1}}

\newcommand{\Tab}[1]{Tab.~\ref{#1}}

\newcommand{\folg}[3]{#1_{#2}, \ldots, #1_{#3}}

\begin{document}

\title*{Universal neural field computation}
\titlerunning{Universal neural field computation}

\author{
Peter beim Graben
    \and
Roland Potthast
}
\institute{
Peter beim Graben \\
    Department of German Studies and Linguistics, \\
    Bernstein Center for Computational Neuroscience Berlin, \\
    Humboldt-Universit\"at zu Berlin, Germany \\
        \and
Roland Potthast \\
    Department of Mathematics and Statistics, \\
    University of Reading, UK and \\
		Deutscher Wetterdienst, Frankfurter Str. 135, \\
		63067 Offenbach, Germany
}

\maketitle

\abstract{
Turing machines and G\"odel numbers are important pillars of the theory of computation. Thus, any computational architecture needs to show how it could relate to Turing machines and how stable implementations of Turing computation are possible. In this chapter, we implement universal Turing computation in a neural field environment. To this end, we employ the canonical symbologram representation of a Turing machine obtained from a G\"odel encoding of its symbolic repertoire and generalized shifts. The resulting nonlinear dynamical automaton (NDA) is a piecewise affine-linear map acting on the unit square that is partitioned into rectangular domains. Instead of looking at point dynamics in phase space, we then consider functional dynamics of probability distributions functions (p.d.f.s) over phase space. This is generally described by a Frobenius-Perron integral transformation that can be regarded as a neural field equation over the unit square as feature space of a dynamic field theory (DFT). Solving the Frobenius-Perron equation yields that uniform p.d.f.s with rectangular support are mapped onto uniform p.d.f.s with rectangular support, again. We call the resulting representation \emph{dynamic field automaton}.
}



\section{Introduction}
\label{sec:PG:intro}

Studying the computational capabilities of neurodynamical systems has commenced with the groundbreaking 1943 article of McCulloch and Pitts \cite{McCullochPitts43} on networks of idealized two-state neurons that essentially behave as logic gates. Because nowadays computers are nothing else than large-scale networks of logic gates, it is clear that computers can in principle be build up by neural networks of McCulloch-Pitts units. This has also been demonstrated by a number of theoretical studies reviewed in \cite{Sontag95}. However, even the most powerful modern workstation is, from a mathematical point of view, only a finite state machine due to its rather huge, though limited memory, while a universal computer, formally codified as a Turing machine \cite{HopcroftUllman79, Turing37}, possesses an unbounded memory tape.

Using continuous-state units with a sigmoidal activation function, Siegelmann and Sontag \cite{SiegelmannSontag95} were able to prove that a universal Turing machine can be implemented by a recurrent neural network of about 900 units, most of them describing the machine's control states, while the tape is essentially represented by a plane spanned by the activations of just two units. The same construction, employing a G\"odel code \cite{Goedel31, Hofstadter79} for the tape symbols, has been previously used by Moore \cite{Moore90, Moore91a} for proving the equivalence of nonlinear dynamical automata and Turing machines. Along a different vain, deploying sequential cascaded networks, Pollack \cite{Pollack91} and later Moore \cite{Moore98} and Tabor \cite{Tabor00a, TaborChoSzkudlarek13} introduced and further generalized dynamical automata as nonautonomous dynamical systems. An even further generalization of dynamical automata, where the tape space becomes represented by a function space, lead Moore and Crutchfield \cite{MooreCrutchfield00} to the concept of a quantum automaton (see \cite{GrabenPotthast09a} for a review and some unified treatment of these different approaches).

Quite remarkably, another paper from McCulloch and Pitts published in 1947 \cite{PittsCulloch47} already set up the groundwork for such functional representations in continuous neural systems. Here, those pioneers investigated distributed neural activation over cortical or subcortical maps representing visual or auditory feature spaces. These neural fields are copied onto many layers, each transforming the field according to a particular member of a symmetry group. For these, a number of field functionals is applied to yield a group invariant that serves for subsequent pattern detection. As early as in this publication, we already find all necessary ingredients for a \emph{Dynamic Field Architecture}: a layered system of neural fields defined over appropriate feature spaces \cite{ErlhagenSchoner02, Schoner02} (see also the chapter of Lins and Sch\"oner in this volume).

We begin this chapter with a general exposition of dynamic field architectures in \Sec{sec:PG:princomp} where we illustrate how variables and structured data types on the one hand and algorithms and sequential processes on the other hand can be implemented in such environments. In \Sec{sec:PG:dfa} we review known facts about nonlinear dynamical automata and introduce dynamic field automata from a different perspective. The chapter is concluded with a short discussion about universal computation in neural fields.


\section{Principles of Universal Computation}
\label{sec:PG:princomp}

As already suggested by McCulloch and Pitts \cite{PittsCulloch47} in 1947, a neural, or likewise, \emph{dynamic field architecture} is a layered system of dynamic neural fields $u_i(x, t) \in \mathbb{R}$ where $1 \le i \le n$ ($i, n \in \mathbb{N}$) indicates the layer, $x \in D$ denotes spatial position in a suitable $d$-dimensional \emph{feature space} $D \subset \mathbb{R}^d$ and $t \in \mathbb{R}_0^+$ time. Usually, the fields obey the Amari neural field equation \cite{Amari77b}
\begin{equation}\label{eq:PG:amari}
    \tau_i \frac{\partial u_i(x, t)}{\partial t} = - u_i(x, t) + h(x) + \sum_{j = 1}^n \int_D w_{ij}(x, y) f(u_j(y, t)) \, \D y + p_i(x, t) \:,
\end{equation}
where $\tau_i$ is a characteristic time scale of the $i$-th layer, $h(x)$ the unique resting activity, $w_{ij}(x, y)$ the synaptic weight kernel for a connection to site $x$ in layer $i$ from site $y$ in layer $j$,
\begin{equation}\label{eq:PG:sigmoid}
    f(u) = \frac{1}{1 + \E^{-\beta(u - \theta)}}
\end{equation}
is a sigmoidal activation function with gain $\beta$ and threshold $\theta$, and $p_i(x, t)$ external input delivered to site $x$ in layer $i$ at time $t$. Note, that a two-layered architecture could be conveniently described by a one-layered complex neural field $z(x, t) = u_1(x, t) + \I u_2(x, t)$  as used in \cite{GrabenEA08b, GrabenPotthast09a, GrabenPotthast12a}.

Commonly, \Eq{eq:PG:amari} is often simplified in the literature by assuming one universal time constant $\tau$, by setting $h=0$ and by replacing $p_i$ through appropriate initial, $u_i(x, 0)$, and boundary conditions, $u_i(\partial D, t)$. With these simplifications, we have to solve the Amari equation
\begin{equation}\label{eq:PG:amari2}
    \tau \frac{\partial u_i(x, t)}{\partial t} = - u_i(x, t) + \sum_{j = 1}^n \int_D w_{ij}(x, y) f(u_j(y, t)) \, \D y
\end{equation}
for initial condition, $u_i(x, 0)$, stating a computational task. Solving that task is achieved through a transient dynamics of \Eq{eq:PG:amari2} that eventually settles down either in an attractor state or in a distinguished terminal state $U_i(x, T)$, after elapsed time $T$. Mapping one state into another, which again leads to a transition to a third state and so on, we will see how the field dynamics can be interpreted as a kind of universal computation, carried out by a program encoded in the particular kernels $w_{ij}(x, y)$, which are in general heterogeneous, i.e. they are not pure convolution kernels: $w_{ij}(x, y) \ne w_{ij}(||x- y||)$ \cite{GrabenHuttSB, JirsaKelso00}.


\subsection{Variables and data types }
\label{sec:PG:data}

How can \emph{variables} be realized in a neural field environment? At the hardware-level of conventional digital computers, variables are sequences of bytes stored in random access memory (RAM). Since a byte is a word of eight bits and since nowadays RAM chips have about 2 to 8 gigabytes, the computer's memory appears as an approximately $8 \times 4 \cdot 10^9$ binary matrix, similar to an image of black-white pixels. It seems plausible to regard this RAM image as a discretized neural field, such that the value of $u(x, t)$ at $x \in D$ could be interpreted as a particular instantiation of a variable. However, this is not tenable for at least two reasons. First, such variables would be highly volatile as bits might change after every processing cycle. Second, the required function space would be a ``mathematical monster'' containing highly discontinuous functions that are not admitted for the dynamical law \pref{eq:PG:amari2}. Therefore, variables have to be differently introduced into neural field computers by assuring temporal stability and spatial smoothness.

We first discuss the second point. Possible solutions to the neural field equation \pref{eq:PG:amari2} must belong to appropriately chosen function spaces that allow the storage and retrieval of variables through binding and unbinding operations. A variable is stored in the neural field by binding its value to an address and its value is retrieved by the corresponding unbinding procedure. These operations have been described in the framework of \emph{Vector Symbolic Architectures} \cite{Gayler06, Smolensky90} and applied to dynamic neural fields by beim Graben and Potthast \cite{GrabenPotthast09a} through a three-tier top-down approach, called \emph{Dynamic Cognitive Modeling}, where variables are regarded as instantiations of data types of arbitrary complexity, ranging from primitive data types such as characters, integers, or floating numbers, over arrays (strings, vectors and matrices) of those primitives, up to structures and objects that allow the representation of lists, frames or trees. These data types are in a first step decomposed into \emph{filler/role bindings} \cite{Smolensky90} which are sets of ordered pairs of sets of ordered pairs etc, of so-called fillers and roles. Simple fillers are primitives whereas roles address the appearance of a primitive in a complex data type. These addresses could be, e.g., array indices or tree positions. Such filler/role bindings can recursively serve as complex fillers bound to other roles. In a second step, fillers and roles are identified with particular basis functions over suitable feature spaces while the binding is realized through functional tensor products with subsequent compression (e.g. by means of convolution products) \cite{Plate95, Smolensky06}.

Since the complete variable allocation of a conventional digital computer can be viewed as an instantiation of only one complex data type, namely an array containing every variable at a particular address, it is possible to map a total variable allocation onto a compressed tensor product in function space of a dynamic field architecture. Assuming that the field $u$ encodes such an allocation, a new variable $\varphi$ in its functional tensor product representation is stored by binding it first to a new address $\psi$, yielding $\varphi \otimes \psi$ and second by superimposing it with the current allocation, i.e. $u + \varphi \otimes \psi$. Accordingly, the value of $\varphi$ is retrieved through an unbinding $\langle \psi^+, u \rangle$ where $\psi^+$ is the adjoint of the address $\psi$ where $\varphi$ is bound to. These operations require further underlying structure of the employed function spaces that are therefore chosen as Banach or Hilbert spaces where either adjoint or bi-orthogonal basis functions are available (see \cite{GrabenGerth12, GrabenEA08b, GrabenPotthast09a, GrabenPotthast12a, PotthastGraben09a} for examples).

The first problem was the volatility of neural fields. This has been resolved using attractor neural networks \cite{Hertz95, Hopfield84} where variables are stabilized as asymptotically stable fixed points. Since a fixed point is defined through $\dot{u}_i(x, t) = 0$, the field obeys the equation
\begin{equation}\label{eq:PG:amarifix}
     u_i(x, t)  = \sum_{j = 1}^n \int_D w_{ij}(x, y) f(u_j(y, t)) \, \D y \:.
\end{equation}
This is achieved by means of a particularly chosen kernel $w_{ii}(||x - y||)$ with local excitation and global inhibition, often called lateral inhibition kernels \cite{ErlhagenSchoner02, Schoner02}.


\subsection{Algorithms and sequential processes}
\label{sec:PG:proc}

Conventional computers run programs that dynamically change variables. Programs perform algorithms that are sequences of instructions, including operations upon variables, decisions, loops, etc. From a mathematical point of view, an algorithm is an element of an abstract algebra that has a representation as an operator on the space of variable allocations, which is well-known as \emph{denotational semantics} in computer science \cite{Tennent76}. The algebra product is the concatenation of instructions being preserved in the representation which is thereby an algebra homomorphism \cite{GrabenGerth12, GrabenPotthast09a}. Concatenating instructions or composing operators takes place step-by-step in discrete time. Neural field dynamics, as governed by \Eq{eq:PG:amari2}, however requires continuous time. How can sequential algorithms be incorporated into the continuum of temporal evolution?

Looking first at conventional digital computers again suggests a possible solution: computers are clocked. Variables remain stable during a clock cycle and gating enables instructions to access variable space. A similar approach has recently been introduced to dynamic field architectures by Sandamirskaya and Sch\"oner \cite{SandamirskayaSchoner08, SandamirskayaSchoner10a}. Here a sequence of neural field activities is stored in a stack of layers, each stabilized by a lateral inhibition kernel. One state is destabilized by a gating signal provided by a condition-of-satisfaction mechanism playing the role of the ``clock'' in this account. Afterwards, the decaying pattern in one layer, excites the next relevant field in a subsequent layer.

Another solution, already outlined in our dynamic cognitive modeling framework \cite{GrabenPotthast09a}, identifies the intermediate results of a computation with saddle fields that are connected their respective stable and unstable manifolds to form stable heteroclinic sequences \cite{RabinovichEA08, AfraimovichZhigulinRabinovich04, GrabenHuttSB}. We have utilized this approach in \cite{GrabenPotthast12a} for a dynamic field model of syntactic language processing. Moreover, the chosen model of winnerless competition among neural populations \cite{FukaiTanaka97} allowed us to explicitly construct the synaptic weight kernel from the filler/role binding of syntactic phrase structure trees \cite{GrabenPotthast12a}.


\section{Dynamic Field Automata}
\label{sec:PG:dfa}

In this section we elaborate our recent proposal on \emph{dynamic field automata} \cite{GrabenPotthast12b} by crucially restricting function spaces to spaces with Haar bases which are piecewise constant fields $u(x, t)$ for $x \in D$, i.e.
\begin{equation}\label{eq:PG:uniform}
    u(x, t) =
        \begin{cases}
        \alpha(t) & \quad:\quad x \in A(t) \\
        0 & \quad:\quad x \notin A(t)
    \end{cases}
\end{equation}
with some time-dependent amplitude $\alpha(t)$ and a possibly time-dependent domain $A(t) \subset D$. Note, that we consider only one-layered neural fields in the sequel for the sake of simplicity.

For such a choice, we first observe that the application of the nonlinear activation function $f$ yields another piecewise constant function over $D$:
\begin{equation}\label{eq:PG:uni1}
    f(u(x, t)) =
        \begin{cases}
        f(\alpha(t)) & \quad:\quad x \in A(t) \\
        f(0) & \quad:\quad x \notin A(t) \:,
    \end{cases}
\end{equation}
which can be significantly simplified by the choice $f(0) = 0$, that holds, e.g., for the linear identity $f = \mathrm{id}$, for the Heaviside step function $f = \Theta$ or for the hyperbolic tangens, $f = \tanh$.

With this simplification, the input integral of the neural field becomes
\begin{equation}\label{eq:PG:nftint}
    \int_D w(x, y) f(u(y, t)) \; \D y = \int_{A(t)} w(x, y) f(\alpha(t)) \; \D y = f(\alpha(t)) \int_{A(t)} w(x, y) \; \D y \:.
\end{equation}

When we additionally restrict ourselves to piecewise constant kernels as well, the last integral becomes
\begin{equation}\label{eq:PG:nftint2}
    \int_{A(t)} w(x, y) \; \D y = w |A(t)|
\end{equation}
with $w$ as constant kernel value and $|A(t)|$ the measure (i.e. the volume) of the domain $A(t)$. Inserting \pref{eq:PG:nftint} and \pref{eq:PG:nftint2} into the fixed point equation \pref{eq:PG:amarifix} yields
\begin{equation}\label{eq:PG:fixhaar}
    u_0 = |A(t)| \cdot w \cdot f(u_0)
\end{equation}
for the fixed point $u_0$. Next, we carry out a linear stability analysis
\begin{eqnarray}\label{eq:PG:lambda}
\dot{u} & = & - u + |A(t)| w f(u) \\
& = & -(u_0 + (u - u_0)) + |A(t)| w \Big( f(u_0) + f'(u_0)\cdot( u - u_0) \Big) +
        O(|u-u_0|^2) \nonumber \\
& = & \Big(- 1 + |A(t)| w f'(u_0)\Big) \cdot( u - u_0) + O(|u-u_0|^2) \:. \nonumber
\end{eqnarray}
Thus, we conclude that if $|A(t)| w f'(u_0) < 1$, then $\dot{u} < 0$ for $u > u_0$ and conversely, $\dot{u} > 0$ for $u < u_0$ in a neighborhood of $u_0$, such that $u_0$ is an asymptotically stable fixed point of the neural field equation.

\begin{figure}[H]
 \includegraphics[width=\textwidth]{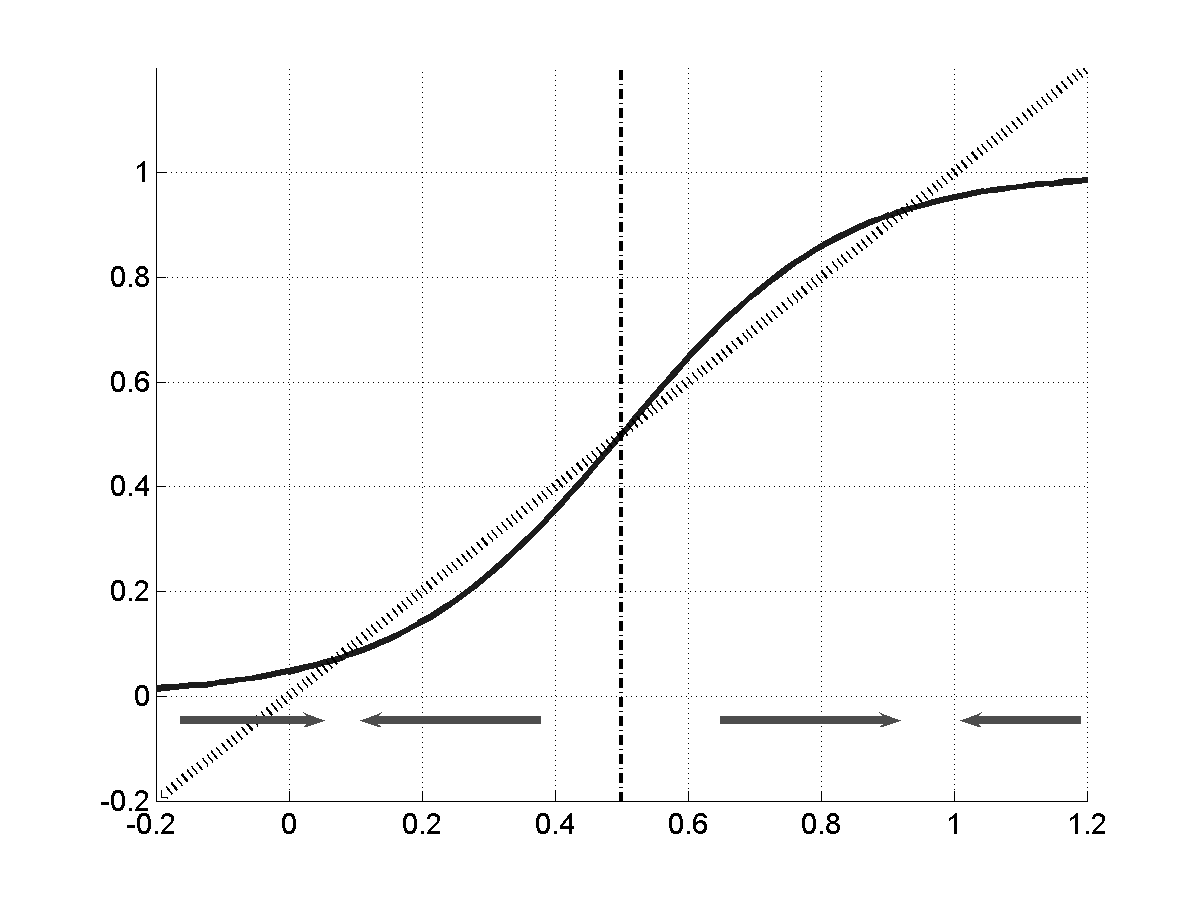}
\caption{
 \label{fig:PG:stabil} Stability of piecewise constant neural field $u_0(x, t)$ over a domain $A \subset D$. Shown are the sigmoidal activation function $f(u)$ (solid) and $u$ (dotted) for comparison. The axis here are given in terms of
absolute numbers without unit as employed in equations (\ref{eq:PG:sigmoid}) or (\ref{eq:PG:amari2}).
 }
\end{figure}

Of course, linear stability analysis is a standard tool to investigate the behavior of dynamic fields around fixed points. For our particular situation it is visualized in \Fig{fig:PG:stabil}. When the solid curve displaying  $|A(t)| w f(u)$ is above $u$ (the dotted curve), then the dynamics \pref{eq:PG:lambda} leads to an increase of $u$, indicated by the arrows pointing to the right. In the case where $|A(t)| w f(u)<u$, a decrease of $u$ is obtained from \pref{eq:PG:lambda}. This is indicated by the arrows pointing to the left. When we have three points where the curves coincide, \Fig{fig:PG:stabil} shows that the setting leads to two stable fixed-points of the dynamics. When the activity field $u(x)$ reaches any value close to these fixed points, the dynamics leads them to the fixed-point values $u_0$.

\subsection{Turing machines}
\label{sec:PG:Turing}

For the construction of dynamic field automata through neural fields we next consider discrete time that might be supplied by some clock mechanism. This requires the stabilization of the fields \pref{eq:PG:uniform} within one clock cycle which can be achieved by self-excitation with a nonlinear activation function $f$ as described in \pref{eq:PG:lambda}, leading to stable excitations as long as we do not include inhibitive elements, where a subsequent state would inhibit those states which were previously excited.

\begin{figure}[H]
\centering
 \subfigure[]{\includegraphics[width=0.45\textwidth]{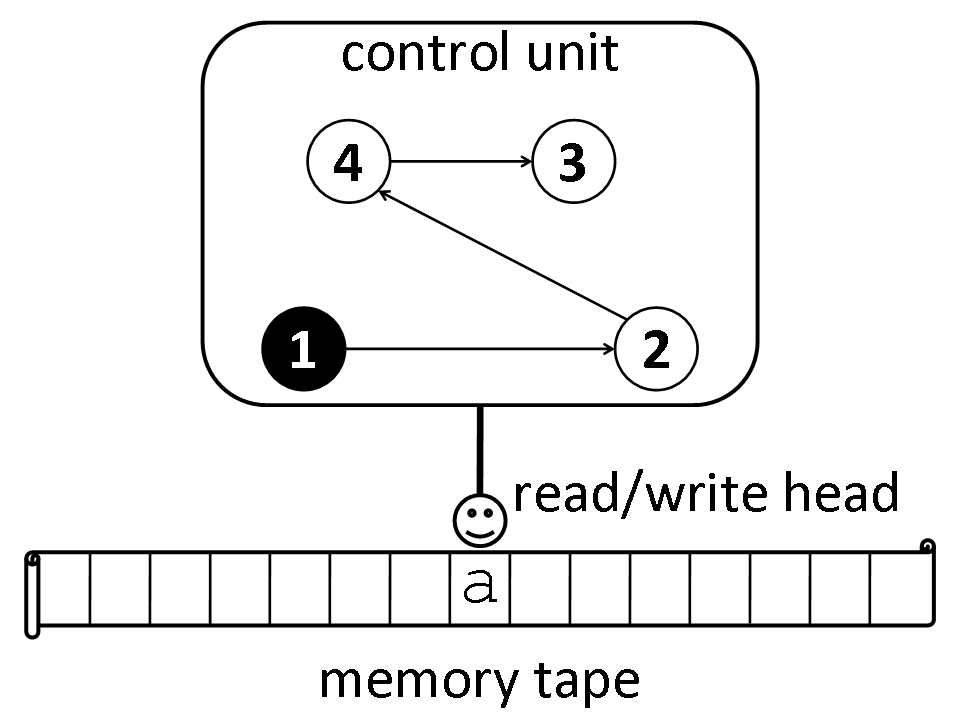}}
 \subfigure[]{\includegraphics[width=0.45\textwidth]{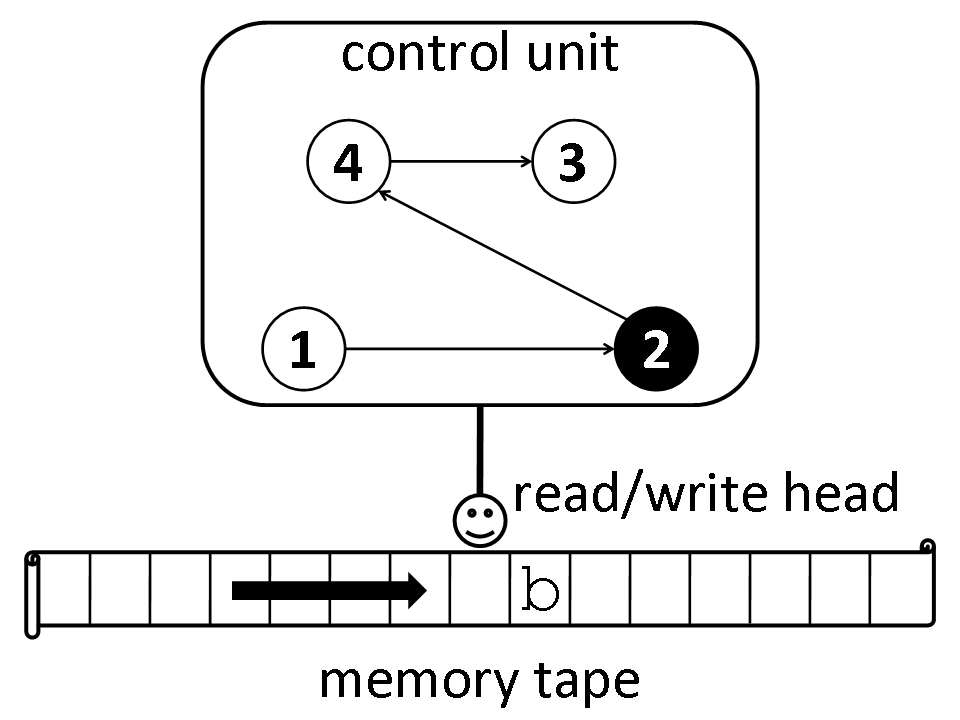}}
\caption{\label{fig:PG:TuringM} Example state transition from (a) to (b) of a Turing machine with $\delta(1, \mathtt{a}) = (2, \mathtt{b}, R) $}.
\end{figure}

Next we briefly summarize some concepts from theoretical computer science \cite{HopcroftUllman79, GrabenPotthast09a, Turing37}. A \emph{Turing machine} is formally defined as a 7-tuple  $M_{TM} = ( Q, \mathbf{N}, \mathbf{T}, \delta, q_0, b, F )$, where $Q$ is a finite set of machine control states, $\mathbf{N}$ is another finite set of tape symbols, containing a distinguished ``blank'' symbol $b$, $\mathbf{T} \subset \mathbf{N} \setminus \{ b \}$ is input alphabet, and
\begin{equation}
 \label{eq:PG:TuringDelta}
 \delta: Q \times \mathbf{N} \to Q \times \mathbf{N} \times \{ L, R \}
\end{equation}
is a partial state transition function, the so-called ``machine table'', determining the action of the machine when $q \in Q$ is the current state at time $t$ and $a \in \mathbf{N}$ is the current symbol at the memory tape under the read/write head. The machine moves then into another state $q' \in Q $ at time $t + 1$ replacing the symbol $a$ by another symbol $a' \in \mathbf{N}$ and shifting the tape either one place to the left (``$L$'') or to the right (``$R$''). Figure \ref{fig:PG:TuringM} illustrates such a state transition. Finally, $q_0 \in Q$ is a distinguished initial state and $F \subset Q$ is a set of ``halting states'' that are assumed when a computation terminates \cite{HopcroftUllman79}.

A Turing machine becomes a time- and state-discrete dynamical system by introducing \emph{state descriptions}, which are triples
\begin{equation}\label{eq:PG:TMstatedesc}
 s = (\alpha, q, \beta)
\end{equation}
where $\alpha, \beta \in \mathbf{N}^*$ are strings of tape symbols to the left and to the right from the head, respectively.  $\mathbf{N}^*$ contains all strings of tape symbols from $\mathbf{N}$ of arbitrary, yet finite, length, delimited by blank symbols $b$. Then, the transition function can be extended to state descriptions by
\begin{equation}\label{eq:PG:TuringDeltaStar}
 \delta^* : S \to S \:,
\end{equation}
where $S = \mathbf{N}^* \times Q \times \mathbf{N}^*$ now plays the role of a phase space of a discrete dynamical system. The set of tape symbols and machine states then becomes a larger alphabet $\mathbf{A} = \mathbf{N} \cup Q$.

Moreover, state descriptions can be conveniently expressed by means of bi-infinite ``dotted sequences''
\begin{equation}\label{eq:PG:symseq}
    s = \ldots a_{i_{-3}} a_{i_{-2}} a_{i_{-1}} . a_{i_{0}} a_{i_{1}} a_{i_{2}} \ldots
\end{equation}
with symbols $a_{i_k} \in \mathbf{A}$. In \Eq{eq:PG:symseq} the dot denotes the observation time $t = 0$ such that the symbol left to the dot, $a_{i_{-1}}$, displays the current state, dissecting the string $s$ into two one-sided infinite strings $s = (\alpha', \beta)$ with $\alpha' = a_{i_{-1}} a_{i_{-2}} a_{i_{-3}} \ldots$ as the left-hand part in reversed order and $\beta = a_{i_{0}} a_{i_{1}} a_{i_{2}} \ldots$

In symbolic dynamics, a cylinder set \cite{McMillan53} is a subset of the space $\mathbf{A}^\mathbb{Z}$ of bi-infinite sequences from an alphabet $\mathbf{A}$ that agree in a particular building block of length $n \in \mathbb{N}$ from a particular instance of time $t \in \mathbb{Z}$, i.e.
\begin{equation}\label{eq:PG:cylinder}
    C(n, t) = [\folg a {i_1} {i_n} ] = 
    \{ s \in \mathbf{A}^{\mathbb{Z}} \,| \, s_{t + k - 1} = 
            a_{i_k} , \quad k = 1, \dots, n \}
\end{equation}
is called $n$-cylinder at time $t \in \mathbb{Z}$. When now $t < 0, n > |t| + 1$ the cylinder contains the dotted word $w = s_{-1} . s_0$ and can therefore be decomposed into a pair of cylinders $(C'(|t|, t), C(|t| + n - 1, 0))$ where $C'$ denotes reversed order of the defining strings again. 

A \emph{generalized shift} \cite{Moore90, Moore91a} emulating a Turing machine is a pair $M_{GS} = (\mathbf{A}^\mathbb{Z}, \Psi)$ where $\mathbf{A}^\mathbb{Z}$ is the space of dotted sequences with $s \in \mathbf{A}^\mathbb{Z}$ and $\Psi : \mathbf{A}^\mathbb{Z} \to \mathbf{A}^\mathbb{Z}$ is given as
\begin{equation}\label{eq:PG:genshift1}
    \Psi(s) = \sigma^{F(s)}(s \oplus G(s))
\end{equation}
with
\begin{eqnarray}\label{eq:PG:genshift}
  \label{eq:genshift2} F: \mathbf{A}^\mathbb{Z} &\to& \mathbb{Z} \\
  \label{eq:genshift3} G: \mathbf{A}^\mathbb{Z} &\to& \mathbf{A}^e \:,
\end{eqnarray}
where $\sigma : \mathbf{A}^\mathbb{Z} \to \mathbf{A}^\mathbb{Z}$ is the left-shift known from symbolic dynamics \cite{LindMarcus95}, $F(s) = l$ dictates a number of shifts to the right ($l < 0$), to the left ($l > 0$) or no shift at all ($l = 0$), $G(s)$ is a word $w'$ of length $e \in \mathbb{N}$ in the domain of effect (DoE) replacing the content $w \in \mathbf{A}^d$, which is a word of length $d \in \mathbb{N}$, in the domain of dependence (DoD) of $s$, and $s \oplus G(s)$ denotes this replacement function.

A generalized shift becomes a Turing machine by interpreting $a_{i_{-1}}$ as the current control state $q$ and $a_{i_{0}}$ as the tape symbol currently underneath the head. Then the remainder of $\alpha$ is the tape left to the head and the remainder of $\beta$ is the tape right to the head. The DoD is the word $w = a_{i_{-1}} . a_{i_{0}}$ of length $d = 2$.

As an instructive example we consider a toy model of syntactic language processing. In order to process a sentence such as ``the dog chased the cat'', linguists often derive a \emph{context-free grammar} (CFG) from a phrase structure tree (see \cite{GrabenGerthVasishth08} for a more detailed example). In our case such a CFG could consist of \emph{rewriting rules}
\begin{align}
    \mathtt{S} &\to \mathtt{NP \ VP} \label{eq:PG:cfg1}\\
    \mathtt{VP} &\to \mathtt{V \ NP} \label{eq:PG:cfg2}\\
    \mathtt{NP} &\to \mathtt{the \ dog} \label{eq:PG:cfg3}\\
    \mathtt{V} &\to \mathtt{chased} \label{eq:PG:cfg4}\\
    \mathtt{NP} &\to \mathtt{the \ cat} \label{eq:PG:cfg5}
\end{align}
where the left-hand side always presents a nonterminal symbol to be expanded into a string of nonterminal and terminal symbols at the right-hand side. Omitting the lexical rules (\ref{eq:PG:cfg3} -- \ref{eq:PG:cfg5}), we regard the symbols $\mathtt{NP}, \mathtt{V}$, denoting ``noun phrase'' and ``verb'', respectively, as terminals and the symbols $\mathtt{S}$ (``sentence'') and $\mathtt{VP}$ (``verbal phrase'') as nonterminals.

A generalized shift processing this grammar is then prescribed by the mappings
\begin{equation}\label{eq:PG:gs}
\begin{array}{l @{\:\mapsto \:}l }
    \mathtt{S} . a & \mathtt{VP \ NP} . a \\
    \mathtt{VP} . a & \mathtt{NP \ V} . a \\
    Z . a & \epsilon . \epsilon
\end{array}
\end{equation}
where the left-hand side of the tape is now called ``stack'' and the right-hand side ``input''.
In \pref{eq:PG:gs}
$Z \in \mathbf{N}$ denotes an arbitrary stack symbol whereas
$a \in\mathbf{T}$ stands for an input symbol.
The empty word is indicated by $\epsilon$. Note the reversed order for the stack left of the dot. The first two operations in \pref{eq:PG:gs} are \emph{predictions} according to a rule of the CFG while the last one is an attachment of input material with already predicted material, to be understood as a matching step.

With this machine table, a \emph{parse} of the sentence ``the dog chased the cat'' ($\texttt{NP V NP}$) is then obtained in \Tab{tab:PG:parse}.
\begin{table}[H]
  \centering
\begin{tabular}{cr@{ . }ll}
  \hline
  time & \multicolumn{2}{c}{state} & operation \\
  \hline
  0 & \texttt{S} & \texttt{NP V NP} & predict \pref{eq:PG:cfg1} \\
  1 & \texttt{VP NP} & \texttt{NP V NP} & attach \\
  2 & \texttt{VP} & \texttt{V NP} & predict \pref{eq:PG:cfg2}  \\
  3 & \texttt{NP V} & \texttt{V NP} & attach  \\
  4 & \texttt{NP} & \texttt{NP} & attach  \\
  5 & $\epsilon$ & $\epsilon$  & accept \\
  \hline
\end{tabular}
\caption{\label{tab:PG:parse} Sequence of state transitions of the generalized shift processing the well-formed string ``the dog chased the cat'' ($\texttt{NP V NP}$). The operations are indicated as follows: ``predict (X)'' means prediction according to rule (X) of the context-free grammar; attach means cancelation of successfully predicted terminals both from stack and input; and ``accept'' means acceptance of the string as being well-formed.}
\end{table}


\subsection{Nonlinear dynamical automata}
\label{sec:PG:nda}

Applying a G\"odel encoding \cite{Goedel31, GrabenPotthast09a, Hofstadter79}
\begin{eqnarray}\label{eq:PG:goedel}
    x &=& \psi(\alpha') := \sum_{k = 1}^\infty \psi(a_{i_{-k}}) b_L^{-k} \\
    y &=& \psi(\beta) := \sum_{k = 0}^\infty \psi(a_{i_k}) b_R^{-k - 1} \nonumber
\end{eqnarray}
to the pair $s = (\alpha', \beta)$ from the Turing machine state description \pref{eq:PG:symseq} where $\psi(a_j) \in \mathbb{N}_0$ is an integer G\"odel number for symbol $a_j \in \mathbf{A}$ and $b_L, b_R \in \mathbb{N}$ are the numbers of symbols that could appear either in $\alpha'$ or in $\beta$, respectively, yields the so-called symbol plane or \emph{symbologram representation} $\vec{x} =(x, y)^T$ of $s$ in the unit square $X$ \cite{CvitanovicGunaratneProcaccia88, KennelBuhl03}.

The symbologram representation of a generalized shift is a \emph{nonlinear dynamical automaton} (NDA) \cite{GrabenPotthast09a, GrabenGerthVasishth08, GrabenJurishEA04}) which is a triple $M_{NDA} = (X, \mathcal{P}, \Phi)$ where $(X, \Phi)$ is a time-discrete dynamical system with phase space $X = [0, 1]^2 \subset \mathbb{R}^2$, the unit square, and flow $\Phi : X \to X$. $\mathcal{P} = \{ D_\nu | \nu = (i, j), 1 \le i \le m, 1 \le j \le n,  m, n \in \mathbb{N} \}$ is a rectangular partition of $X$ into pairwise disjoint sets, $D_\nu \cap D_\mu = \emptyset$ for $\nu \ne \mu$, covering the whole phase space $X = \bigcup_\nu D_\nu$, such that $D_\nu = I_i \times J_j$ with real intervals $I_i, J_j \subset [0, 1]$ for each bi-index $\nu = (i, j)$. The cells $D_\nu$ are the domains of the branches of $\Phi$ which is a piecewise affine-linear map
\begin{equation}\label{eq:PG:ndamap}
    \Phi(\vec{x}) =
    \begin{pmatrix}
      a^{\nu}_x \\
      a^{\nu}_y
    \end{pmatrix} +
       \begin{pmatrix}
      \lambda^{\nu}_x & 0 \\
      0 & \lambda^{\nu}_y
    \end{pmatrix} \cdot
    \begin{pmatrix}
      x \\
      y
    \end{pmatrix} \:,
\end{equation}
when $\vec{x} = (x, y)^T \in D_\nu$. The vectors $(a^{\nu}_x, a^{\nu}_y )^T \in \mathbb{R}^2$ characterize parallel translations, while the matrix coefficients $\lambda^{\nu}_x, \lambda^{\nu}_y \in \mathbb{R}_0^+$ mediate either stretchings ($\lambda > 1$), squeezings  ($\lambda < 1$), or identities ($\lambda = 1$) along the $x$- and $y$-axes, respectively. Here, the letters $x$ and $\nu$ at $a^{\nu}_{x}$ or
$\lambda_{x}^{\nu}$ indicate the dependence of the coefficients on $x$ and the index of the particular cylinder set $D_{\nu}$ under consideration.

Hence, the NDA's dynamics, obtained by iterating an orbit $\{ \vec{x}_t \in X | t \in \mathbb{N}_0 \}$ from initial condition $\vec{x}_0$ through
\begin{equation}\label{eq:PG:ndadyn}
    \vec{x}_{t + 1} = \Phi(\vec{x}_t)
\end{equation}
describes a symbolic computation by means of a generalized shift \cite{Moore90, Moore91a} when subjected to the coarse-graining $\mathcal{P}$.

The domains of dependence and effect (DoD and DoE) of an NDA, respectively, are obtained as images of cylinder sets under the G\"odel encoding \pref{eq:PG:goedel}. Each cylinder possesses a lower and an upper bound, given by the G\"odel numbers 0 and $b_L - 1$ or $b_R - 1$, respectively. Thus,
\begin{eqnarray*}
  \inf(\psi(C'(|t|, t))) &=& \psi(\folg a {i_{|t|}} {i_1}) \\
  \sup(\psi(C'(|t|, t))) &=& \psi(\folg a {i_{|t|}} {i_1}) + b_L^{-|t|} \\
  \inf(\psi(C(|t| + n - 1, 0))) &=& \psi(\folg a {i_{|t| + 1}} {i_n}) \\
  \sup(\psi(C(|t| + n - 1, 0))) &=& \psi(\folg a {i_{|t| + 1}} {i_n}) + b_R^{-|t| - n + 1} \:,
\end{eqnarray*}
where the suprema have been evaluated by means of geometric series \cite{GrabenJurishEA04}. Thereby, each part cylinder $C$ is mapped onto a real interval $[\inf(C), \sup(C)] \subset [0, 1]$ and the complete cylinder $C(n, t)$ onto the Cartesian product of intervals $R = I \times J \subset [0, 1]^2$, i.e. onto a rectangle in unit square. In particular, the empty cylinder, corresponding to the empty tape $\epsilon . \epsilon$ is represented by the complete phase space $X = [0, 1]^2$.

Fixing the prefixes of both part cylinders and allowing for random symbolic continuation beyond the defining building blocks, results in a cloud of randomly scattered points across a rectangle $R$ in the symbologram \cite{GrabenGerthVasishth08}. These rectangles are consistent with the symbol processing dynamics of the NDA, while individual points $\vec{x} \in [0, 1]^2$ no longer have an immediate symbolic interpretation. Therefore, we refer to arbitrary rectangles $R \in [0, 1]^2$ as to NDA macrostates, distinguishing them from NDA microstates $\vec{x}$ of the underlying dynamical system.

\begin{figure}[H]
\centering
 \subfigure[]{\includegraphics[width=0.45\textwidth]{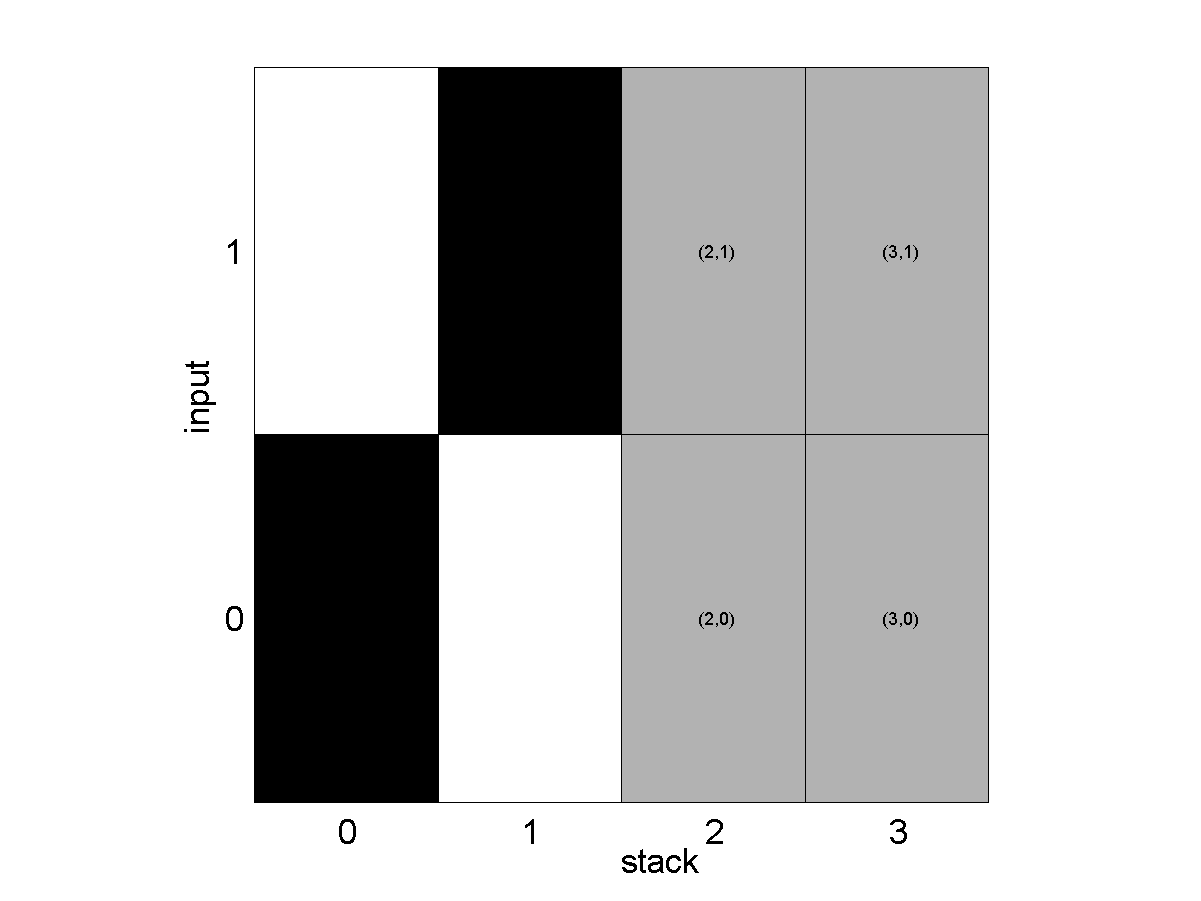}}
 \subfigure[]{\includegraphics[width=0.45\textwidth]{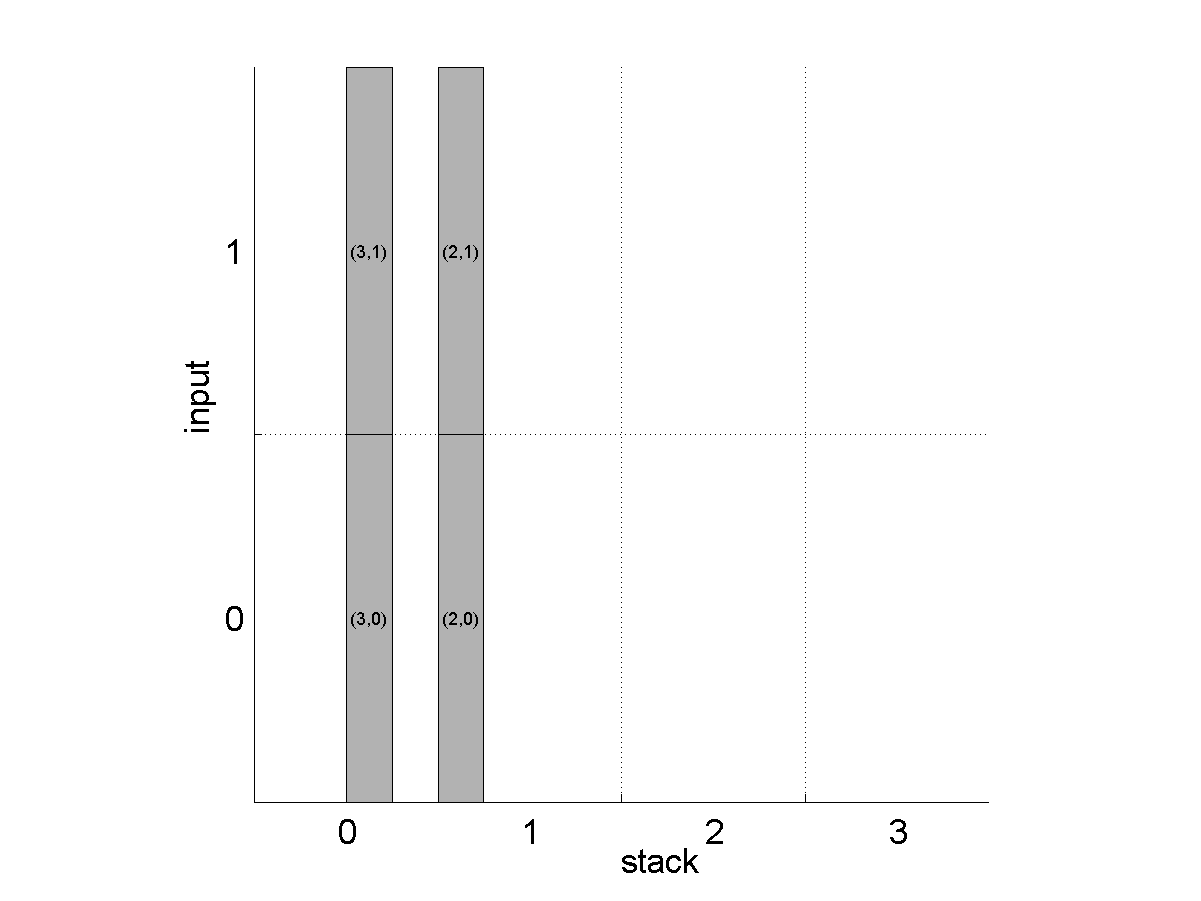}}
\caption{\label{fig:PG:nda} Symbologram of the NDA processing the string ``the dog chased the cat'' ($\texttt{NP V NP}$).  (a) Domains of dependence (DoD) of actions: identity (white), predict (gray), and attach (black). (b) Domains of effect (DoE): images of prediction (gray), black rectangles from (a) are mapped onto the whole unit square during attachment}.
\end{figure}

Coming back to our language example, we create an NDA from an arbitrary G\"odel encoding. Choosing
\begin{align}\label{eq:PG:gcode}
 \Psi(\mathtt{NP})    &= 0 \\
 \Psi(\mathtt{V})     &= 1 \\
 \Psi(\mathtt{VP})    &= 2 \\
 \Psi(\mathtt{S})     &= 3 \\
\end{align}
we have $b_L = 4$ stack symbols and $b_R = 2$ input symbols. Thus, the symbologram is partitioned into eight rectangles. Figure \ref{fig:PG:nda} displays the resulting (a) DoD and (b) DoE.

\subsection{Neural field computation}
\label{sec:PG:nftcomp}

Next we replace the NDA point dynamics in phase space by functional dynamics in Banach space. Instead of iterating clouds of randomly prepared initial conditions according to a deterministic dynamics, we consider the deterministic dynamics of probability measures over phase space. This higher level of description that goes back to Koopman et al. \cite{Koopman31, KoopmanVonNeumann32} has recently been revitalized for dynamical systems theory \cite{BudivsicMohrMezic12}.

The starting point for this approach is the conservation of probability as expressed by the Frobenius-Perron equation \cite{Ott93}
\begin{equation}\label{eq:PG:froper}
    \rho(\vec{x}, t) = \int_X \delta(\vec{x} - \Phi^{t - t'}(\vec{x}')) \rho(\vec{x}', t') \D \vec{x}' \:,
\end{equation}
where $\rho(\vec{x}, t)$ denotes a probability density function over the phase space $X$ at time $t$ of a dynamical system,  $\Phi^t : X \to X$ refers to either a continuous-time ($t \in \mathbb{R}_0^+$) or discrete-time ($t \in \mathbb{N}_0$) flow and the integral over the delta function expresses the probability summation of alternative trajectories all leading into the same state $\vec{x}$ at time $t$.

In the case of an NDA, the flow is discrete and piecewise affine-linear on the domains $D_\nu$ as given by \Eq{eq:PG:ndamap}. As initial probability distribution densities $\rho(\vec{x}, 0)$ we consider uniform distributions with rectangular support $R_0 \subset X$, corresponding to an initial NDA macrostate,
\begin{equation}\label{eq:PG:iniuni}
    u(\vec{x}, 0) = \frac{1}{|R_0|} \chi_{R_0}(\vec{x}) \:,
\end{equation}
where
\begin{equation}\label{eq:PG:charfn}
   \chi_{A}(\vec{x}) = \begin{cases}
                                        0 & \quad:\quad \vec{x} \notin A \\
                                        1 & \quad:\quad \vec{x} \in A
                                    \end{cases}
\end{equation}
is the characteristic function for a set $A \subset X$. A crucial requirement for these distributions is that they must be consistent with the partition $\mathcal{P}$ of the NDA, i.e. there must be a bi-index $\nu = (i, j)$ such that the support $R_0 \subset D_\nu$.

Inserting \pref{eq:PG:iniuni} into the Frobenius-Perron equation \pref{eq:PG:froper} yields for one iteration
\begin{equation}\label{eq:PG:froper2}
    u(\vec{x}, t + 1) = \int_X \delta(\vec{x} - \Phi(\vec{x}')) u(\vec{x}', t) \D \vec{x}' \:.
\end{equation}

In order to evaluate \pref{eq:PG:froper2}, we first use the product decomposition of the involved functions:
\begin{equation}\label{eq:PG:decouni1}
    u(\vec{x}, 0) = u_x(x, 0) u_y(y, 0)
\end{equation}
with
\begin{eqnarray}
    \label{eq:PG:decounix}
    u_x(x, 0) &=& \frac{1}{|I_0|} \chi_{I_0}(x) \\
    \label{eq:PG:decouniy}
    u_y(y, 0) &=& \frac{1}{|J_0|} \chi_{J_0}(y)
\end{eqnarray}
and
\begin{equation}\label{eq:PG:decdelta}
  \delta(\vec{x} - \Phi(\vec{x}')) = \delta(x - \Phi_x(\vec{x}')) \delta(y - \Phi_y(\vec{x}'))  \:,
\end{equation}
where the intervals $I_0, J_0$ are the projections of $R_0$ onto $x$- and $y$-axes, respectively. Correspondingly, $\Phi_x$ and $\Phi_y$ are the projections of $\Phi$ onto $x$- and $y$-axes, respectively. These are obtained from \pref{eq:PG:ndamap} as
\begin{eqnarray}
    \label{eq:PG:mapx}
    \Phi_x(\vec{x}') &=& a^{\nu}_x + \lambda^{\nu}_x x' \\
    \label{eq:PG:mapy}
    \Phi_y(\vec{x}') &=& a^{\nu}_y + \lambda^{\nu}_y y' \:.
\end{eqnarray}
Using this factorization, the Frobenius-Perron equation \pref{eq:PG:froper2} separates into
\begin{eqnarray}
    \label{eq:PG:fpx}
    u_x(x, t + 1) &=& \int_{[0, 1]} \delta(x - a^{\nu}_x - \lambda^{\nu}_x x') u_x(x', t) \D x' \\
    \label{eq:PG:fpy}
    u_y(y, t + 1) &=& \int_{[0, 1]} \delta(y - a^{\nu}_y - \lambda^{\nu}_y y') u_y(y', t) \D y'
\end{eqnarray}

Next, we evaluate the delta functions according to the well-known lemma
\begin{equation}\label{eq:PG:evadelta}
 \delta(f(x)) = \sum_{l : \text{simple zeros}} |f'(x_l)|^{-1} \delta(x - x_l) \:,
\end{equation}
where $f'(x_l)$ indicates the first derivative of $f$ in $x_l$. \Eq{eq:PG:evadelta} yields for the $x$-axis
\begin{equation}\label{eq:PG:zeros}
    x_\nu = \frac{x - a^{\nu}_x}{\lambda^{\nu}_x} \:,
\end{equation}
i.e. one zero for each $\nu$-branch, and hence
\begin{equation}\label{eq:PG:slope}
    |f'(x_\nu')| = \lambda^{\nu}_x \:.
\end{equation}
Inserting \pref{eq:PG:evadelta}, \pref{eq:PG:zeros} and \pref{eq:PG:slope} into \pref{eq:PG:fpx}, gives
\begin{eqnarray*}
   u_x(x, t + 1) &=& \sum_\nu \int_{[0, 1]} \frac{1}{\lambda^{\nu}_x} \delta\left( x' - \frac{x - a^{\nu}_x}{\lambda^{\nu}_x} \right) u_x(x', t) \D x' \\
    &=& \sum_\nu \frac{1}{\lambda^{\nu}_x} u_x\left( \frac{x - a^{\nu}_x}{\lambda^{\nu}_x}, t \right)
\end{eqnarray*}

Next, we take into account that the distributions must be consistent with the NDA's partition. Therefore, for given $\vec{x} \in D_\nu$ there is only one branch of $\Phi$ contributing a simple zero to the sum above. Hence,
\begin{equation}\label{eq:PG:uniter}
    u_x(x, t + 1) =
    \sum_\nu \frac{1}{\lambda^{\nu}_x} u_x\left( \frac{x - a^{\nu}_x}{\lambda^{\nu}_x}, t \right) =
    \frac{1}{\lambda^{\nu}_x} u_x\left( \frac{x - a^{\nu}_x}{\lambda^{\nu}_x}, t \right) \:.
\end{equation}

Our main finding is now that the evolution of uniform p.d.f.s with rectangular support according to the NDA dynamics \Eq{eq:PG:froper2} is governed by
\begin{equation}\label{eq:PG:fpuniform}
    u(\vec{x}, t) = \frac{1}{|\Phi^t(R_0)|} \chi_{\Phi^t(R_0)}(\vec{x}) \:,
\end{equation}
i.e. uniform distributions with rectangular support are mapped onto uniform distributions with rectangular support \cite{GrabenPotthast12b}.

For the proof we first insert the initial uniform density distribution \pref{eq:PG:iniuni} for $t = 0$ into \Eq{eq:PG:uniter}, to obtain by virtue of \pref{eq:PG:decounix}
\[
        u_x(x, 1) = \frac{1}{\lambda^{\nu}_x} u_x\left( \frac{x - a^{\nu}_x}{\lambda^{\nu}_x}, 0 \right) =
        \frac{1}{\lambda^{\nu}_x} \frac{1}{|I_0|} \chi_{I_0}\left( \frac{x - a^{\nu}_x}{\lambda^{\nu}_x} \right) \:.
\]
Deploying \pref{eq:PG:charfn} yields
\[
    \chi_{I_0}\left( \frac{x - a^{\nu}_x}{\lambda^{\nu}_x} \right) =
    \begin{cases}
        0 & \quad:\quad \frac{x - a^{\nu}_x}{\lambda^{\nu}_x}  \notin I_0 \\
        1 & \quad:\quad \frac{x - a^{\nu}_x}{\lambda^{\nu}_x} \in I_0 \:.
    \end{cases}
\]
Let now $I_0 = [p_0, q_0] \subset [0, 1]$ we get
\begin{eqnarray*}
    && \frac{x - a^{\nu}_x}{\lambda^{\nu}_x} \in I_0 \\
    &\Longleftrightarrow&
    p_0 \le \frac{x - a^{\nu}_x}{\lambda^{\nu}_x} \le q_0 \\
    &\Longleftrightarrow&
    \lambda^{\nu}_x p_0 \le x - a^{\nu}_x \le \lambda^{\nu}_x q_0 \\
    &\Longleftrightarrow&
    a^{\nu}_x + \lambda^{\nu}_x p_0 \le x \le a^{\nu}_x + \lambda^{\nu}_x q_0 \\
    &\Longleftrightarrow&
    \Phi_x(p_0) \le x  \le \Phi_x(q_0) \\
    &\Longleftrightarrow&
    x  \in \Phi_x(I_0) \:,
\end{eqnarray*}
where we made use of \pref{eq:PG:mapx}.
Moreover, we have
\[
 \lambda^{\nu}_x |I_0| = \lambda^{\nu}_x (q_0 - p_0) = q_1 - p_1 = |I_1|
\]
with $I_1 = [p_1, q_1] = \Phi_x(I_0)$. Therefore,
\[
        u_x(x, 1) = \frac{1}{|I_1|} \chi_{I_1}(x) \:.
\]

The same argumentation applies to the $y$-axis, such that we eventually obtain
\begin{equation}\label{eq:PG:uniter2}
    u(\vec{x}, 1) = \frac{1}{|R_1|} \chi_{R_1}(\vec{x}) \:,
\end{equation}
with $R_1 = \Phi(R_0) $ the image of the initial rectangle $R_0 \subset X$. Thus, the image of a uniform density function with rectangular support is a uniform density function with rectangular support again.

Next, assume \pref{eq:PG:fpuniform} is valid for some $t \in \mathbb{N}$. Then it is obvious that \pref{eq:PG:fpuniform} also holds for $t + 1$ by inserting the $x$-projection of \pref{eq:PG:fpuniform} into \pref{eq:PG:uniter} using \pref{eq:PG:decounix}, again. Then, the same calculation as above applies when every occurrence of $0$ is replaced by $t$ and every occurrence of $1$ is replaced by $t + 1$. By means of this construction we have implemented an NDA by a dynamically evolving field. Therefore, we call this representation dynamic field automaton (DFA).

The Frobenius-Perron equation \pref{eq:PG:froper2} can be regarded as a time-discretized Amari dynamic neural field equation \pref{eq:PG:amari2}. Discretizing time according to Euler's rule with increment $\Delta t = \tau$ where $\tau$ is the time constant of the Amari equation \pref{eq:PG:amari2} yields
\begin{eqnarray*}
    \tau \frac{u(\vec{x}, t + \tau) - u(\vec{x}, t)}{\tau} + u(\vec{x}, t) &=&
  \int_D w(\vec{x}, \vec{y}) f(u(\vec{y}, t)) \; \D \vec{y} \\
  u(\vec{x}, t + \tau) &=& \int_D w(\vec{x}, \vec{y}) f(u(\vec{y}, t)) \; \D \vec{y} \:.
\end{eqnarray*}
For $\tau = 1$ and $f(u) = u$ the Amari equation becomes the Frobenius-Perron equation \pref{eq:PG:froper2} when we set
\begin{equation}\label{eq:PG:nftkernel}
    w(\vec{x}, \vec{y}) = \delta(\vec{x} - \Phi(\vec{y}))
\end{equation}
where $\Phi$ is the NDA mapping from \Eq{eq:PG:ndadyn}. This is the general solution of the kernel construction problem \cite{PotthastGraben09a, GrabenPotthast09a}. Note that $\Phi$ is not injective, i.e.\ for fixed $\vec{x}$ the kernel is a sum of delta functions coding the influence from different parts of the space $X = [0,1]^2$.

\begin{figure}[H]
\centering
 \includegraphics[width=\textwidth]{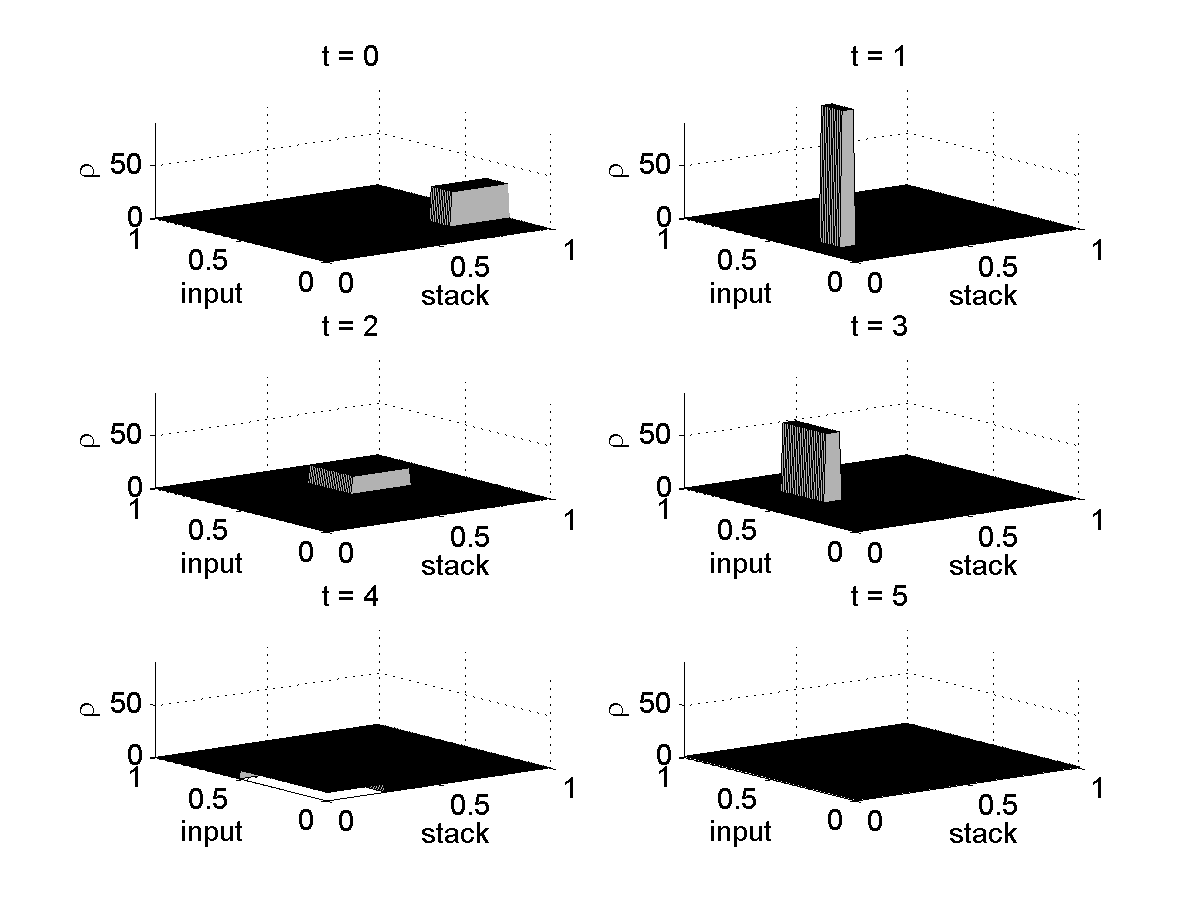}
\caption{\label{fig:PG:dfa} Dynamic field automaton for processing the string ``the dog chased the cat'' ($\texttt{NP V NP}$) according to \Tab{tab:PG:parse}. The NDA states become rectangular supports of uniform distributions which are mapped onto uniform distributions with rectangular supports during discrete temporal evolution.}
\end{figure}

Finally we carry out the whole construction for our language example. This yields the field dynamics depicted in \Fig{fig:PG:dfa}.

\section{Discussion}
\label{sec:PG:disc}

Turing machines and G\"odel numbers are important pillars of the theory of computation \cite{HopcroftUllman79, SpencerTanayEA13}. Thus, any computational architecture needs to show how it could relate to Turing machines and in what way stable implementations of Turing computation is possible. In this chapter, we addressed the question how universal Turing computation could be implemented in a neural field environment as described by its easiest possible form, the Amari field equation (\ref{eq:PG:amari}). To this end, we employed the canonical symbologram representation \cite{CvitanovicGunaratneProcaccia88, KennelBuhl03} of the machine tape as the unit square, resulting from a G\"odel encoding of sequences of states.

The action of the Turing machine on a state description is given by a state flow on the unit square which led to a Frobenius-Perron equation \pref{eq:PG:froper} for the evolution of uniform probability densities. We have implemented this equation in the neural field space by a piecewise affine-linear kernel geometry on the unit square which can be expressed naturally within a neural field framework. We also showed that stability of states and dynamics both in time as well as its encoding for finite programs is achieved by the approach.

However, our construction essentially relied upon discretized time that could be provided by some clock mechanism. The crucial problem of stabilizing states within every clock cycle could be principally solved by established methods from dynamic field architectures. In such a time-continuous extension, an excited state, represented by a rectangle in one layer, will only excite a subsequent state, represented by another rectangle in another layer when a condition-of-satisfaction is met \cite{SandamirskayaSchoner08, SandamirskayaSchoner10a}. Otherwise rectangular states would remain stabilized as described by \Eq{eq:PG:lambda}. All these problems provide promising prospects for future research.


\begin{acknowledgement}
We thank Slawomir Nasuto and Serafim Rodrigues for helpful comments improving this chapter. This research was supported by a DFG Heisenberg fellowship awarded to PbG (GR 3711/1-2).
\end{acknowledgement}

\nocite{Arbib95}


\begin{thebibliography}{10}
\providecommand{\url}[1]{{#1}}
\providecommand{\urlprefix}{URL }
\expandafter\ifx\csname urlstyle\endcsname\relax
  \providecommand{\doi}[1]{DOI~\discretionary{}{}{}#1}\else
  \providecommand{\doi}{DOI~\discretionary{}{}{}\begingroup
  \urlstyle{rm}\Url}\fi

\bibitem{AfraimovichZhigulinRabinovich04}
Afraimovich, V.S., Zhigulin, V.P., Rabinovich, M.I.: On the origin of
  reproducible sequential activity in neural circuits.
\newblock Chaos \textbf{14}(4), 1123 -- 1129 (2004).

\bibitem{Amari77b}
Amari, S.I.: Dynamics of pattern formation in lateral-inhibition type neural
  fields.
\newblock Biological Cybernetics \textbf{27}, 77 -- 87 (1977)

\bibitem{Arbib95}
Arbib, M.A. (ed.): The Handbook of Brain Theory and Neural Networks, 1st edn.
\newblock MIT Press, Cambridge (MA) (1995)

\bibitem{BudivsicMohrMezic12}
Budi\v{s}i\'{c}, M., Mohr, R., Mezi\'{c}, I.: {Applied Koopmanism}.
\newblock Chaos \textbf{22}(4), 047510 (2012).

\bibitem{CvitanovicGunaratneProcaccia88}
Cvitanovi\'c, P., Gunaratne, G.H., Procaccia, I.: Topological and metric
  properties of {H\'enon-type} strange attractors.
\newblock Physical Reviews A \textbf{38}(3), 1503 -- 1520 (1988)

\bibitem{ErlhagenSchoner02}
Erlhagen, W., Sch\"oner, G.: Dynamic field theory of movement preparation.
\newblock Psychological Review \textbf{109}(3), 545 -- 572 (2002).

\bibitem{FukaiTanaka97}
Fukai, T., Tanaka, S.: A simple neural network exhibiting selective activation
  of neuronal ensembles: from winner-take-all to winners-share-all.
\newblock Neural Computation \textbf{9}(1), 77 -- 97 (1997).

\bibitem{Gayler06}
Gayler, R.W.: Vector symbolic architectures are a viable alternative for
  {Jackendoff's} challenges.
\newblock Behavioral and Brain Sciences \textbf{29}, 78 -- 79 (2006).

\bibitem{Goedel31}
G\"odel, K.: {\"Uber} formal unentscheidbare {S\"atze} der {\em {principia}
  mathematica} und verwandter {Systeme} {I}.
\newblock Monatshefte f\"ur Mathematik und Physik \textbf{38}, 173 -- 198
  (1931)

\bibitem{GrabenGerth12}
beim Graben, P., Gerth, S.: Geometric representations for minimalist grammars.
\newblock Journal of Logic, Language and Information \textbf{21}(4), 393 -- 432
  (2012).

\bibitem{GrabenGerthVasishth08}
beim Graben, P., Gerth, S., Vasishth, S.: Towards dynamical system models of
  language-related brain potentials.
\newblock Cognitive Neurodynamics \textbf{2}(3), 229 -- 255 (2008).

\bibitem{GrabenHuttSB}
beim Graben, P., Hutt, A.: Attractor and saddle node dynamics in heterogeneous
  neural fields. Submitted.

\bibitem{GrabenJurishEA04}
beim Graben, P., Jurish, B., Saddy, D., Frisch, S.: Language processing by
  dynamical systems.
\newblock International Journal of Bifurcation and Chaos \textbf{14}(2), 599 --
  621 (2004)

\bibitem{GrabenEA08b}
beim Graben, P., Pinotsis, D., Saddy, D., Potthast, R.: Language processing
  with dynamic fields.
\newblock Cognitive Neurodynamics \textbf{2}(2), 79 -- 88 (2008).

\bibitem{GrabenPotthast09a}
beim Graben, P., Potthast, R.: Inverse problems in dynamic cognitive modeling.
\newblock Chaos \textbf{19}(1), 015103 (2009).

\bibitem{GrabenPotthast12a}
beim Graben, P., Potthast, R.: A dynamic field account to language-related
  brain potentials.
\newblock In: M.~Rabinovich, K.~Friston, P.~Varona (eds.) Principles of Brain
  Dynamics: Global State Interactions, chap.~5, pp. 93 -- 112. MIT Press,
  Cambridge (MA) (2012)

\bibitem{GrabenPotthast12b}
beim Graben, P., Potthast, R.: Implementing turing machines in dynamic field
  architectures.
\newblock In: M.~Bishop, Y.J. Erden (eds.) Proceedings of AISB12 World Congress
  2012 - Alan Turing 2012, vol. 5th AISB Symposium on Computing and Philosophy:
  Computing, Philosophy and the Question of Bio-Machine Hybrids, pp. 36 -- 40
  (2012).
\newblock \urlprefix\url{http://arxiv.org/abs/1204.5462}

\bibitem{Hertz95}
Hertz, J.: Computing with attractors.
\newblock In: Arbib  \cite{Arbib95}, pp. 230 -- 234

\bibitem{Hofstadter79}
Hofstadter, D.R.: G\"odel, Escher, Bach: an Eternal Golden Braid.
\newblock Basic Books, New York (NY) (1979)

\bibitem{HopcroftUllman79}
Hopcroft, J.E., Ullman, J.D.: Introduction to Automata Theory, Languages, and
  Computation.
\newblock Addison--Wesley, Menlo Park, California (1979)

\bibitem{Hopfield84}
Hopfield, J.J.: Neurons with graded response have collective computational
  properties like those of two-state neurons.
\newblock Proceedings of the National Academy of Sciences of the U.S.A.
  \textbf{81}(10), 3088 -- 3092 (1984).

\bibitem{JirsaKelso00}
Jirsa, V.K., Kelso, J.A.S.: Spatiotemporal pattern formation in neural systems
  with heterogeneous connection toplogies.
\newblock Physical Reviews E \textbf{62}(6), 8462 -- 8465 (2000)

\bibitem{KennelBuhl03}
Kennel, M.B., Buhl, M.: Estimating good discrete partitions from observed data:
  Symbolic false nearest neighbors.
\newblock Physical Review Letters \textbf{91}(8), 084,102 (2003)

\bibitem{Koopman31}
Koopman, B.O.: Hamiltonian systems and transformations in {H}ilbert space.
\newblock Proceedings of the National Academy of Sciences of the U.S.A.
  \textbf{17}, 315 -- 318 (1931)

\bibitem{KoopmanVonNeumann32}
Koopman, B.O., von Neumann, J.: Dynamical systems of continuous spectra.
\newblock Proceedings of the National Academy of Sciences of the U.S.A.
  \textbf{18}, 255 -- 262 (1932)

\bibitem{LindMarcus95}
Lind, D., Marcus, B.: An Introduction to Symbolic Dynamics and Coding.
\newblock Cambridge University Press, Cambridge (UK) (1995).

\bibitem{McCullochPitts43}
McCulloch, W.S., Pitts, W.: A logical calculus of ideas immanent in nervous
  activity.
\newblock Bulletin of Mathematical Biophysics \textbf{5}, 115 -- 133 (1943)

\bibitem{McMillan53}
McMillan, B.: The basic theorems of information theory.
\newblock Annals of Mathematical Statistics \textbf{24}, 196 -- 219 (1953)

\bibitem{Moore90}
Moore, C.: Unpredictability and undecidability in dynamical systems.
\newblock Physical Review Letters \textbf{64}(20), 2354 -- 2357 (1990)

\bibitem{Moore91a}
Moore, C.: Generalized shifts: unpredictability and undecidability in dynamical
  systems.
\newblock Nonlinearity \textbf{4}, 199 -- 230 (1991)

\bibitem{Moore98}
Moore, C.: Dynamical recognizers: real-time language recognition by analog
  computers.
\newblock Theoretical Computer Science \textbf{201}, 99 -- 136 (1998)

\bibitem{MooreCrutchfield00}
Moore, C., Crutchfield, J.P.: Quantum automata and quantum grammars.
\newblock Theoretical Computer Science \textbf{237}, 275 -- 306 (2000)

\bibitem{Ott93}
Ott, E.: Chaos in Dynamical Systems.
\newblock Cambridge University Press, New York (1993).

\bibitem{PittsCulloch47}
Pitts, W., McCulloch, W.S.: How we know universals: The perception of auditory
  and visual forms.
\newblock Bulletin of Mathematical Biophysics \textbf{9}, 127 -- 147 (1947)

\bibitem{Plate95}
Plate, T.A.: Holographic reduced representations.
\newblock IEEE Transactions on Neural Networks \textbf{6}(3), 623 -- 641
  (1995).

\bibitem{Pollack91}
Pollack, J.B.: The induction of dynamical recognizers.
\newblock Machine Learning \textbf{7}, 227 -- 252 (1991).
\newblock Also published in \cite{PortvanGelder95}, pp. 283 -- 312.

\bibitem{PortvanGelder95}
Port, R.F., van Gelder, T. (eds.): Mind as Motion: Explorations in the Dynamics
  of Cognition.
\newblock MIT Press, Cambridge (MA) (1995)

\bibitem{PotthastGraben09a}
Potthast, R., beim Graben, P.: Inverse problems in neural field theory.
\newblock SIAM Jounal on Applied Dynamical Systems \textbf{8}(4), 1405 -- 1433
  (2009).

\bibitem{RabinovichEA08}
Rabinovich, M.I., Huerta, R., Varona, P., Afraimovich, V.S.: Transient
  cognitive dynamics, metastability, and decision making.
\newblock PLoS Computational Biology \textbf{4}(5), e1000,072 (2008).

\bibitem{SandamirskayaSchoner08}
Sandamirskaya, Y., Sch\"oner, G.: Dynamic field theory of sequential action: A
  model and its implementation on an embodied agent.
\newblock In: Proceedings of the 7th IEEE International Conference on
  Development and Learning (ICDL), pp. 133 -- 138 (2008).

\bibitem{SandamirskayaSchoner10a}
Sandamirskaya, Y., Sch\"oner, G.: An embodied account of serial order: How
  instabilities drive sequence generation.
\newblock Neural Networks \textbf{23}(10), 1164 -- 1179 (2010).

\bibitem{Schoner02}
Sch\"oner, G.: Neural systems and behavior: Dynamical systems approaches.
\newblock In: N.J. Smelser, P.B. Baltes (eds.) International Encyclopedia of
  the Social \& Behavioral Sciences, pp. 10,571 -- 10,575. Pergamon, Oxford
  (2002)

\bibitem{SiegelmannSontag95}
Siegelmann, H.T., Sontag, E.D.: On the computational power of neural nets.
\newblock Journal of Computer and System Sciences \textbf{50}(1), 132 -- 150
  (1995)

\bibitem{Smolensky90}
Smolensky, P.: Tensor product variable binding and the representation of
  symbolic structures in connectionist systems.
\newblock Artificial Intelligence \textbf{46}(1-2), 159 -- 216 (1990).

\bibitem{Smolensky06}
Smolensky, P.: Harmony in linguistic cognition.
\newblock Cognitive Science \textbf{30}, 779 -- 801 (2006)

\bibitem{Sontag95}
Sontag, E.D.: Automata and neural networks.
\newblock In: Arbib  \cite{Arbib95}, pp. 119 -- 123

\bibitem{SpencerTanayEA13}
Spencer, M.C., Tanay, T., Roesch, E.B., Bishop, J.M., Nasuto, S.J.: Abstract
  platforms of computation.
\newblock AISB 2013 (2013)

\bibitem{Tabor00a}
Tabor, W.: Fractal encoding of context-free grammars in connectionist networks.
\newblock Expert Systems: The International Journal of Knowledge Engineering
  and Neural Networks \textbf{17}(1), 41 -- 56 (2000)

\bibitem{TaborChoSzkudlarek13}
Tabor, W., Cho, P.W., Szkudlarek, E.: Fractal analyis illuminates the form of
  connectionist structural gradualness.
\newblock Topics in Cognitive Science \textbf{5}, 634 -- 667 (2013)

\bibitem{Tennent76}
Tennent, R.D.: The denotational semantics of programming languages.
\newblock Communications of the ACM \textbf{19}(8), 437 -- 453 (1976).

\bibitem{Turing37}
Turing, A.M.: On computable numbers, with an application to the {{\em
  Entscheidungsproblem}}.
\newblock Proceedings of the London Mathematical Society \textbf{2}(42), 230 --
  265 (1937).

\end{thebibliography}

\end{document}